\documentclass[aps,pra,twocolumn,amsmath,amssymb,superscriptaddress,floatfix]{revtex4-2}

\usepackage{graphicx}
\usepackage{bm}
\usepackage[utf8]{inputenc}
\usepackage[T1]{fontenc}
\usepackage{mathptmx}
\usepackage{etoolbox}
\usepackage{units}
\usepackage[colorlinks=true, urlcolor  = blue,linkcolor=blue, citecolor=blue,]{hyperref}
\setcitestyle{super}

\begin{document}
\newcommand{\NV}{\ensuremath{\text{NV}^{\mathrm{-}}}}
\newcommand{\NVo}{\ensuremath{\text{NV}^{\mathrm{0}}}}
\newcommand{\Ex}{\ensuremath{\text{E}_{\mathrm{x}}}}
\newcommand{\A}{\ensuremath{\text{A}_{\mathrm{1}}}}
\newcommand{\msz}{\ensuremath{\text{m}_{\mathrm{s}}=0}}
\newcommand{\mso}{\ensuremath{\text{m}_{\mathrm{s}}=\pm 1}}

\preprint{APS/123-QED}


\title[Cavity-assisted resonance fluorescence from a nitrogen-vacancy center in diamond]{\textbf{Cavity-assisted resonance fluorescence from a nitrogen-vacancy center in diamond}} 
\author{Viktoria Yurgens}
\thanks{These two authors contributed equally to this work}

\author{Yannik Fontana}
\email[]{yanniklaurent.fontana@unibas.ch}
\thanks{These two authors contributed equally to this work}

\author{Andrea Corazza}

\author{Brendan J. Shields}
\altaffiliation[Present address: ]{Quantum Network Technologies,  Boston, Massachusetts 02215, USA}

\author{Patrick Maletinsky}

\author{Richard J. Warburton}
\affiliation{Department of Physics, University of Basel, CH-4056 Basel, Switzerland}

\date{\today}


\begin{abstract}
The nitrogen-vacancy center in diamond, owing to its optically addressable and long-lived electronic spin, is an attractive resource for the generation of remote entangled states. However, the center's low native fraction of coherent photon emission, $\sim$3\%, strongly reduces the achievable spin-photon entanglement rates. Here, we couple a nitrogen-vacancy center with a narrow extrinsically broadened linewidth (\unit[159]{MHz}), hosted in a micron-thin membrane, to the mode of an open optical microcavity. The resulting Purcell factor of $\sim$1.8 increases the fraction of zero-phonon line photons to above 44\%, leading to coherent photon emission rates exceeding four times the state of the art under non-resonant excitation. Bolstered by the enhancement provided by the cavity, we for the first time measure resonance fluorescence without any temporal filtering with $>$10 signal-to-laser background ratio. Our microcavity platform would increase spin-spin entanglement success probabilities by more than an order of magnitude compared to existing implementations. Selective enhancement of the center's zero-phonon transitions could furthermore unlock efficient application of quantum optics techniques such as wave-packet shaping or all-optical spin manipulation.
\end{abstract}

\maketitle


The nitrogen-vacancy center (NV) in diamond in its negatively charged state (\NV) offers a direct link between its long-lived spin ground state~\cite{Bar-Gill2013,Maurer2013b,Abobeih2018} and photons thanks to spin-conserving optical transitions~\cite{Robledo2011,Doherty2013}. Spin manipulation via microwaves and optical spin initialization and readout complete the picture of an attractive solid-state spin qubit~\cite{Robledo2011a}. However, the \NV's small optical dipole moment, the high refractive index of diamond, and a $\sim3\%$ zero-phonon line (ZPL) to phonon side-band (PSB) branching ratio (Debye-Waller factor) preclude the generation of spin-photon entanglement at high rates~\cite{Togan2010,Bernien2013,Pompili2021,Hermans2023}.

A strategy to increase the coherent photon flux involves embedding the \NV\ in an optical cavity. Compared to photonic structures designed to improve the collection efficiency such as solid immersion lenses (SILs)~\cite{Jamali2014,Hadden2010}, the Purcell effect induced by the cavity additionally increases the fraction of coherent photons. Early attempts at using nanophotonic resonators (e.g. photonic crystals) collided with the detrimental impact of nanofabrication on the optical properties of the \NV ~\cite{Barclay2009,Faraon2012,Li2015}. An elegant way to overcome this challenge is to place a minimally processed \NV-containing diamond membrane inside an open microcavity~\cite{Janitz2015,Johnson2015,Riedel2017,Ruf2021}. The microcavity offers efficient mode-matching to external optics, optimal positioning of the emitter, and the ability to tune to and selectively enhance specific optical transitions~\cite{Barbour2011,Albrecht2013, Wang2019, HoyJensen2020,Tomm2021, Flaagan2022,Bayer2023,Deshmukh2023,Zifkin2023,Herrmann2023}.

\begin{figure*}
    \includegraphics[width=1\textwidth]{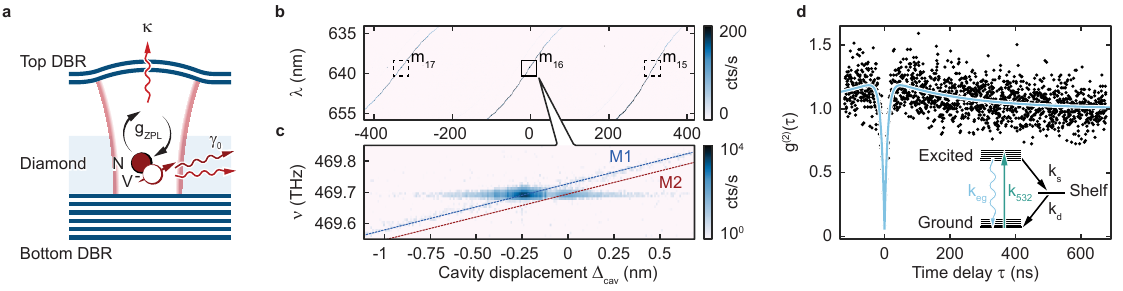}
    \caption{A single \NV\ coupled to a cavity mode. \textbf{a} Schematic of the tunable open microcavity and diamond membrane. For a cavity frequency matching the ZPL of an \NV, a coherent exchange with coupling strength $g_\mathrm{ZPL}$ takes place between the \NV\ and the optical mode. Concurrently, the \NV\ decays into the quasi-free-space continuum at a rate $\gamma_0$ while the intra-cavity field leaks out of the cavity at a rate $\kappa$. \textbf{b} Spectral map over modes 17 to 15. Upon non-resonant excitation, background photoluminescence from the diamond reveals the mode structure. \textbf{c} Close to resonant conditions, the \NV\ selectively couples to the non-degenerate cavity modes M1 and M2 (m$_16$). \textbf{d} Second-order autocorrelation performed on resonance with mode M1 confirms the isolation of a single \NV. The data (including background) is fitted with a three-level rate equations model (inset).}
    \label{fig1}
\end{figure*}

A spin-photon interface based on a single \NV\ coupled to a microcavity presents certain challenges: success hinges on simultaneously retaining low cavity losses and a narrow \NV\ linewidth. So far, attempts have been partially foiled either by non-optimal cavity performance or by excessive fabrication-induced broadening of the NV ZPL linewidth~\cite{Riedel2017,Ruf2021}. While at room temperature, a cavity finesse -- inversely proportional to the round trip losses -- of at most 17\,000~\cite{Janitz2015} has been achieved, the necessary transition to cryogenic temperatures ($<\unit[10]{K}$) needed to prevent phonon-induced dephasing generally leads to reduced cavity performance, with a recent implementation yielding a finesse of 12\,000~\cite{Pallmann2023}, albeit for a dense and inhomogeneously broadened ensemble. Until now, the various degrees of Purcell-enhancement of a single \NV\ emission mostly amounted to values well below unity~\cite{Johnson2015,Ruf2021}. Successful attempts at obtaining significant Purcell factors have fallen short in efficiently extracting the emitted photons, due to intra-cavity losses dominating the optical decay process~\cite{Riedel2017}.

\begin{figure}
    \includegraphics[width=1\columnwidth]{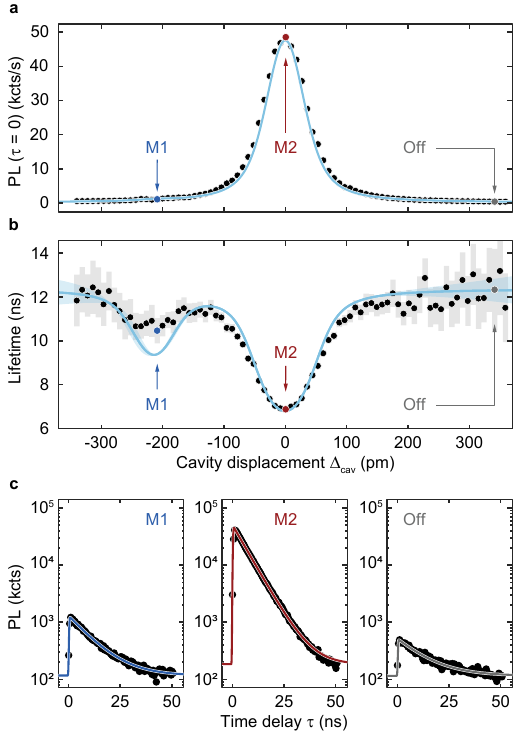}
    \caption{Purcell-enhancement under pulsed non-resonant excitation. \textbf{a} ZPL emission rate at zero-delay (full black circles) for a collection co-polarized with mode M2. \textbf{b} Extracted single-exponential decay times (full circles). The coupling of the \NV\ to either cavity mode yields a lifetime reduction at the respective cavity displacement values. The fit in panels a and b (light blue line) accounts for both modes as well as for residual cavity vibrations. Shaded areas represent 95\% confidence intervals on fits and 1.96 standard deviation on data points. \textbf{c} ZPL temporal decays for cavity displacement values corresponding to mode M1 (navy), mode M2 (burgundy), and off-resonance with either modes (grey).}
    \label{fig2}
\end{figure}

In addition, maintaining high single-photon purity and high indistinguishability requires that the relevant \NV\ transitions are excited resonantly, raising the challenge of filtering scattered photons from the laser background. So far, robust simultaneous resonant excitation and collection of the ZPL has relied mainly on SILs to increase the collection efficiency and reduce the laser background, backed by polarization-based suppression and systematic temporal filtering of the ZPL photons~\cite{Togan2010, Bernien2013, Ruf2021}. While temporal filtering offers signal-to-laser background ratio (SLR) of up to $10^3$ ~\cite{Bernien2013,Pompili2021,Hermans2022}, it leads to a reduced probability of detecting ZPL photons ($\sim80\%$), bound to plummet if a significant Purcell enhancement reduces the \NV\ lifetime.

In this work, we couple a low-charge-noise \NV\ to the mode of a high-finesse open microcavity at cryogenic temperature. We record Purcell-enhanced ZPL count rates far surpassing current state-of-the-art photonic interfaces based on SILs. Furthermore, polarization-based suppression of the resonant excitation light allows the \NV\ resonance fluorescence (RF) to be detected without relying on temporal filtering, establishing the platform as an efficient photonic interface for quantum applications.

\section*{Results}
\subsection*{Single NV coupling and Purcell Enhancement}

The open microcavity, schematically represented in Fig.\,\ref{fig1}a, consists of a planar bottom mirror and a curved top mirror (radius of curvature $\sim$\unit[14]{\textmu m}). Both mirrors are distributed Bragg reflectors (DBRs) with reflectivities aiming at a design finesse of 11\,614. Excitation and collection is performed through an objective via the top mirror, with transmission $\mathcal{T}_{top}$ $\sim10\cdot \mathcal{T}_{bottom}$. The asymmetry in mirror reflectivities results in a one-sided cavity with a decay rate $\kappa$ and ensures that photons emitted out of the cavity are preferentially collected by the objective. The diamond, a \unit[1.6]{\textmu m}-thick membrane, is bonded to the bottom mirror via van der Waals-forces~\cite{Riedel2017,Janitz2015}. It contains \NV\ created by carbon implantation post-fabrication, a method developed to yield low-noise \NV\ centers in microstructures~\cite{Yurgens2022}. A coupled \NV\ radiates into free-space at a natural rate $\gamma_0$ while concurrently exchanging energy at a rate $g_\mathrm{ZPL}$ with a resonant  cavity mode.

Using the positioning flexibility offered by the open microcavity~\cite{Greuter2014}, we isolate and selectively couple an \NV\ to the cavity optical modes. Under non-resonant continuous-wave excitation with a \unit[532]{nm} wavelength laser, a weak background originating from the diamond unveils the cavity modes (Fig.\,\ref{fig1}b, color scale capped at \unit[150]{cts/s}). At regular displacement intervals corresponding to the free-spectral range, the cavity frequency matches an \NV\ ZPL transition (dashed and solid squares in Fig.\,\ref{fig1}b). For all following experiments, we set the cavity length around values corresponding to mode 16, indicated by the solid square in Fig.\,\ref{fig1}b. A finer scan shows a strong increase in photoluminescence (PL) when resonance conditions are met, and reveals a doublet of peaks, degenerate in emission frequency (Fig.\,\ref{fig1}c). The doublet originates from the \unit[34.4]{GHz}-splitting of the cavity's fundamental transverse mode into two orthogonal, linearly polarized modes (M1 and M2, with M2 chosen as the reference mode, i.e. $\Delta_\mathrm{cav}=0$). Changes in the mode splitting along the membrane and a significant reduction of the mode splitting out of the membrane hints at birefringence induced by the bonded diamond. The estimated corresponding anisotropic stress value (\unit[$\sim 65-75$]{MPa}) in the membrane, while not unrealistic, suggests rather a simultaneous straining of the diamond and the DBR, giving rise to the observed birefringence~\cite{Grimsditch1979,Howell2012,Tomm2021b}. The exact intensity ratio of the M1 and M2 peaks depends on the coupling strength of the \NV\ to either mode, combined with the alignment of the polarization-filtered detection channel. From measurements shown in Fig.\,\ref{fig1}b, we infer that the cavity operates in a mode close to the air-confined condition, a configuration less sensitive to surface scattering losses~\cite{Janitz2015}. The measured finesse $\mathcal{F}$ is 4\,330, corresponding to a quality factor of 42\,930 and a decay rate \unit[$\kappa = 2\pi\cdot 11.0$]{GHz}.

We verify that the detected photons stem from a single \NV\ by performing photon autocorrelation measurements ($g^{(2)}(\tau)$). With the \NV\ on-resonance with mode M1, we correlate the arrival times of the PL photons on two avalanche photodetectors (APDs). The result, normalized to the value at long delay times (\unit[33]{\textmu s}), is shown in Fig.\,\ref{fig2}d. A pronounced antibunching dip of g$^{(2)}(0)=0.04\ll 0.5$ is the clear identifier of a single-photon emitter. The very small g$^{(2)}(0)$ value, considering non-resonant excitation, is a signature of the effect of the cavity: the incoherent background PL is filtered out.

The $g^{(2)}(\tau)$ data is fitted using a three-level rate-equation model~\cite{supp}. The choice is justified as the \NV\ i) is pumped non-resonantly and ii) polarizes into \msz, reducing the complexity of the excited state. The inset of Fig.\,\ref{fig1}d shows the effective states of the model: the \NV\ ground and excited states (\msz), and the shelving state including all possible population-trapping mechanisms. The fit closely follows the data and yields the emission, excitation, shelving and de-shelving rates, respectively: $\{k_{eg}, k_{532}, k_s, k_{d} \} = \{101.2, 2.5, 32.0, 3.8\}$ MHz. The model interprets the bunching behavior at delays larger than zero, explained by the partial shelving of the population into states not participating in the emission process, typically either the \NV\ singlet state, the ionized, neutral NV state (\NVo), or a combination of both. We estimate an NV "on-off" ratio, translatable into an occupation of the triplet states, of $\sim$83\%, matching typical values for non-resonantly driven \NV ~\cite{Doherty2013}. Strikingly, $k_{eg}$, the rate representing the spontaneous emission, is noticeably higher than expected for an \NV\ polarized in \msz\ (\unit[101.2]{MHz} versus $\sim$\unit[81]{MHz}) ~\cite{Robledo2011}.

The key physical principle behind this work, the Purcell effect~\cite{Purcell1946}, results in a shortened radiative lifetime for an emitter that couples to a resonant optical mode. In fact, the high $k_{eg}$ rate is the first quantifiable manifestation of an acceleration of the \NV\ dynamics by the cavity (in this case, by mode M1 by a factor $\sim$1.24). Following, we characterize the Purcell-enhancement in our system by exciting the \NV\ non-resonantly with short pulses of \unit[532]{nm} light (\unit[$\sim$70]{ps}) and monitoring the arrival time of PL photons. The non-resonant pulses guarantee that the NV remains in the \NV\ charge state and is \msz\ polarized. The radiative lifetime for any particular value of $\Delta_\mathrm{cav}$ is extracted by fitting the decay signal with a single exponential, taking a common instrument-related rise-time into account. 

With the measurement apparatus adjusted such that the polarization of the detection channel and mode M2 coincide, the cavity is swept across $\Delta_\mathrm{cav}=0$. The signal rate at delay time $\tau=0$ follows a Lorentzian distribution centered at $\Delta_\mathrm{cav}=0$ with a full-width half-maximum (FWHM) of \unit[80]{pm} (Fig.\,\ref{fig2}a). The lifetimes on the other hand show not one, but two dips (Fig.\,\ref{fig2}b). While the dip at $\Delta_\mathrm{cav}=0$, corresponding to the resonance condition with mode M2, is the most pronounced, another clear lifetime reduction dip can be see around $\Delta_\mathrm{cav}$ = \unit[-210]{pm}, corresponding to the resonance condition with mode M1. The coupling between the \NV\ and M1 opens an additional decay channel which can be observed in the decay dynamics, despite not directly detecting photons leaking from mode M1. For positive cavity displacements $\Delta_\mathrm{cav}$, the extracted lifetimes tend toward the original, uncoupled value $\tau_0 =1/\gamma_0$. Note that the decrease in signal strength at such detunings leads to an increased uncertainty (shaded grey bars). 

Interestingly, the lifetime reduction dip and the corresponding signal peak around $\Delta_\mathrm{cav}=0$ exhibit noticeably different widths (\unit[116]{pm} versus \unit[80]{pm} FWHM). The broader features in the lifetime reduction are well explained by taking into account residual relative motion $\sigma_\mathrm{vib}$ between the mirrors (mechanical noise). We derive a model including the coupling of both cavity modes to a single, linearly polarized \NV\ transition as well as vibrations and fit it to the entire dataset. The model closely reproduces the features of the data and provides insight into the observed broadening of the lifetime dips (light blue line in Fig.\,\ref{fig2}). 

Vibrations add noise to the detuning between the cavity modes and the \NV\ transition. As a reduced detuning leads to both faster decay and higher signal, ``early photons'' are over-represented and the extracted lifetimes appear shorter. The deviations between the model and the data around mode M1 are likely due to a temporary increase in background mechanical noise. The fit further provides a value for the angle of the projection of the optical dipole moment with respect to mode M2: $\theta_\mathrm{cav}$ = \unit[33(4)]{$^\circ$}, the residual mechanical vibration: \unit[$\sigma_\mathrm{vib} =22.0(1)$]{pm-rms}, and the vibration-free Purcell factors for M1 and M2: 1.44(9) and 2.068(5). Finally, we can evaluate the \NV\ lifetime unmodified by the cavity, yielding \unit[$\tau_0=12.35(4)$]{ns}, in excellent agreement with previously reported \NV\ excited-state lifetimes in bulk diamond~\cite{Batalov2008,Robledo2011,Tamarat2006}. 

Fig.\,\ref{fig2}c shows decay traces for three specific cavity detunings: on resonance with cavity mode M1, on resonance with cavity mode M2, and far off-resonance with either mode. The extracted lifetimes are \unit[10.6(6)]{ns} for mode M1 and \unit[6.88(4)]{ns} for mode M2. It is worth noting the consistency between the lifetime measured at resonance with mode M1 and the inverse rate $1/k_{eg}$ extracted from the $g^{(2)(\tau)}$ data (\unit[9.9(7)]{ns}). For the far-off resonance trace, the lifetime is found to be \unit[12.3(2)]{ns}, in agreement with the previously mentioned value of $\tau_0$. The vibration-limited lifetimes for M1 and M2, together with $\tau_0$, are used for all following calculations. 

Denoting the \NV\ total, cavity-modified, decay rate as $\gamma_\mathrm{P}$, the overall Purcell-enhancement can be expressed as 
\begin{equation*} \label{eq:FPtot}
F_\mathrm{P} = \gamma_\mathrm{P}/\gamma_0 = 1+\frac{4g_\mathrm{ZPL}^2}{\kappa\gamma_0}.
\end{equation*}
This definition explicitly includes the emission into the incoherent PSB and the free-space continuum. We extract an overall Purcell-factor of $F_{\mathrm{P},\mathrm{M1}} = 1.15(6)$ and $F_{\mathrm{P},\mathrm{M2}} = 1.79(1)$ for mode M1 and mode M2, respectively. With the probability of spontaneous emission into the cavity mode defined as $\beta = (F_\mathrm{P}-1)/F_\mathrm{P}$, we retrieve $\beta_{\mathrm{M1}} = 13(5)\%$ and $\beta_{\mathrm{M2}} = 44.1(6)\%$.

Since only a fraction of the total \NV\ emission -- the ZPL -- is enhanced by the cavity, a small overall Purcell factor conceals a significant Purcell-enhancement of the ZPL alone, which can be described by
\begin{equation*} \label{eq:FP_ZPL}
    F_\mathrm{P}^{ZPL} = \frac{F_\mathrm{P}-(1-\xi_0)}{\xi_0} 
\end{equation*}
with $\xi_0$ denoting the Debye-Waller factor, here taken as 3\%~\cite{Faraon2012}. We obtain a ZPL Purcell-factor of $F_{\mathrm{P},\mathrm{M1}}^{ZPL} = 6.0(3)$ and $F_{\mathrm{P},\mathrm{M2}}^{ZPL} = 27.3(3)$ for mode M1 and mode M2, respectively. 

Alternatively, the effect of the cavity can be seen as modifying the Debye-Waller factor, dramatically increasing the coherent emission probability. The cavity-enhanced Debye-Waller factor can be written as
\begin{equation*}
    \xi_\mathrm{cav} = \beta+\frac{\xi_0}{F_\mathrm{P}} \approx \beta.
\end{equation*}
Following, when mode M2 is tuned into resonance with the \NV, the branching ratio between ZPL and PSB emission is increased by more than an order of magnitude. In other words, almost half of the \NV\ coherent emission ($\xi_\mathrm{cav} \sim \beta_{\mathrm{M2}} = 44.1(6)\%$), normally spoiled by phonons in the bulk, is restored.

An estimate of the \NV-cavity coupling strength can be extracted from $F_{\mathrm{P},\mathrm{M2}}$, completing the set of cavity quantum electrodynamics parameters $\{g_\mathrm{ZPL}, \kappa, \gamma_0\} = 2\pi \cdot \{167(19), 1.1\cdot10^4, 12.89(4)\}$ MHz. Here, we assume that $\gamma_0$ is given purely by the emitter decay rate off-resonance with the cavity. This approximation neglects dephasing effects, yielding a lower bound. A justification to this simplification is that at low enough temperature and provided that the \NV\ transitions are driven strictly resonantly, Fourier-transformed linewidths are achievable~\cite{Hermans2023}. The obtained values, with $\kappa \gg g_\mathrm{ZPL} \gg \gamma_0$, place the interaction in the weak coupling (fast cavity) regime, but also at the onset of the high-cooperativity regime ($C = F_\mathrm{P}-1 > 1$), for which interactions between the cavity field and the emitter become significant at the single-photon level.

\begin{figure*}
    \includegraphics[width=1\textwidth]{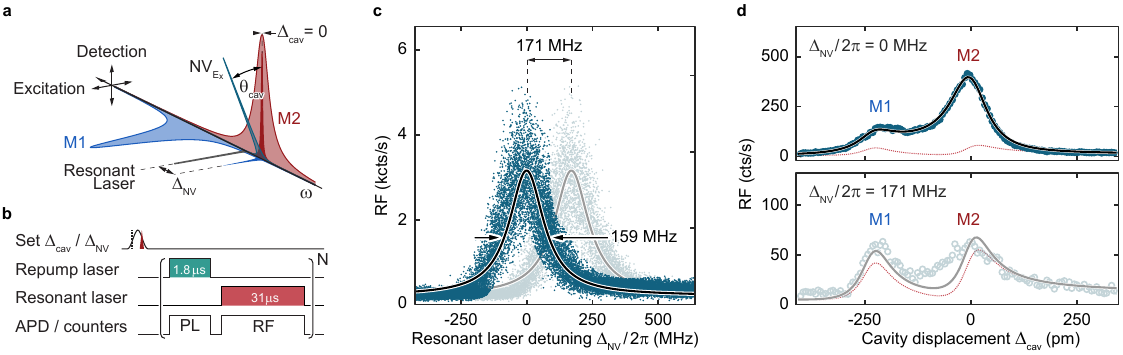}
    \caption{\NV\ resonance fluorescence. \textbf{a} For all RF measurements, the cross-polarized excitation and detection channels are co-polarized with modes M1 (navy) and M2 (burgundy), respectively. Mode M2 is depicted as on-resonance with the \NV\ \Ex\ ZPL transition (teal). Spectral overlap between the modes and the projection of the \Ex\ dipole moment on both modes enable resonant excitation (resonant laser, grey peak) via mode M1 and RF collection via mode M2. \textbf{b} Pulse sequence for RF measurements. ZPL photons are sorted as PL (emitted during the repump pulse) or RF (emitted during the resonant pulse). \textbf{c} RF detected with a resonant power of \unit[1]{nW} at $\Delta_\mathrm{cav}=0$. The teal and blue-gray resonances correspond to the same \Ex\ transition (linewidth \unit[159]{MHz}), Stark-shifted (by \unit[171]{MHz}) by a single charge trapped in the vicinity of the \NV. Zero laser detuning corresponds to a frequency of \unit[469.7]{THz}. \textbf{d} Top: RF signal at $\Delta_\mathrm{NV} = 0$ (\unit[0.1]{nW}). Sweeping the cavity varies the laser-M1 detuning together with the detuning between \Ex\ and both modes. The amplitude ratio of the M1 and M2 peaks is the signature of non-linearity in the response of the \NV. Bottom: Same experiment for $\Delta_{\mathrm{NV}}/2\pi = \unit[171]{MHz}$. The model describing both data sets includes two modes coupling to a single saturable transition (black and grey lines) and includes the laser background (dashed red lines).}
    \label{fig3}
\end{figure*}

\subsection*{Cavity-assisted resonance fluorescence}
A cornerstone of quantum optics is the ability to coherently and resonantly drive a given optical transition. For \NV\ centers, this ability lies at the heart of resonant spin read-out~\cite{Robledo2011a} and spin-photon entanglement generation~\cite{Togan2010,Bernien2013}. A sizeable challenge is to filter the resonantly scattered photons (RF) from the driving laser photons. We obtain laser extinction ratios of \unit[$\sim$50]{dB} by aligning the excitation and detection channels with modes M1 and M2, as illustrated in Fig.\,\ref{fig3}a. The respective projections of the \NV\ dipole moment on modes M1 and M2 allows us to drive the transition (in this case the cycling $\mathrm{m_s} = 0 \leftrightarrow \mathrm{E_x}$ transition, referred to as \Ex\ ~\cite{supp}) via mode M1, while detecting RF via the preferentially coupled mode M2. This configuration is kept for all the following experiments. Once again, the cavity plays a crucial role in amplifying i) the RF signal and ii) the intra-cavity power, enabling us to measure RF with high SLR.

The \NV\ is excited using interleaved non-resonant ``repump'' (\unit[532]{nm}, \unit[1.8]{\textmu s}, \unit[200]{\textmu W} if not specified otherwise) and resonant (\unit[637]{nm}, \unit[31]{\textmu s}) pulses separated by \unit[400]{ns} (Fig.\,\ref{fig3}b). The detected photons are either binned or tagged according to their time of arrival, allowing the identification and monitoring of PL and RF signals. The repump pulses are primarily used to convert \NVo\ to \NV\ and to restore the \msz\ population after shelving into \mso\ by the resonant pulses, but we also use the light generated during this interval to measure PL and conduct periodical checks on the cavity displacement, ensuring drift-free operation.

Maintaining the condition $\Delta_\mathrm{cav}=0$ and sweeping the resonant laser detuning $\Delta_\mathrm{NV}$ (center frequency \unit[$\nu_\mathrm{NV}=469.7$]{THz}), we realize a first unambiguous measurement of RF from a Purcell-enhanced \NV (Fig.\,\ref{fig3}c). Two peaks (teal and blue-grey) can be distinguished, both corresponding to \Ex\ and extrinsically broadened by charge noise generated by the high-energy repump pulses. The doublet is a striking manifestation of resolved charge noise: during a repump pulse, a nearby trap state stochastically captures or releases an electron, leading to different Stark fields and the emergence of two distinct center frequencies for \Ex\ (see Methods and Ref.\,\citenum{supp}). The observation of single-trap loading and unloading highlights the quality of the diamond material and of the NV centers formed via carbon implantation post-fabrication~\cite{Yurgens2022}. For the remainder of this study, the \unit[171(3)]{MHz} blue-shifted state is only used to introduce a fixed detuning.

The extrinsic broadening of \Ex\ is best fitted by a model taking into account the expected Purcell-induced increase of the homogeneous linewidth. As confirmed later, the resonant laser power used in this measurement (\unit[1]{nW}) allows us to neglect power broadening and extract a Lorentzian-distributed linewidth with a FWHM \unit[$\Gamma_\mathrm{ext}/2\pi=159(5)$]{MHz}. This value agrees well with previously measured samples created with the same method~\cite{Yurgens2022} and is instrumental in enabling the measurement of RF in the cavity. A broader linewidth (typically \unit[$>$1]{GHz}), common for microfabricated structures, would result in i) a significant spread of $\tilde{\Delta}_\mathrm{cav}$, the pulse-to-pulse detuning between the transition and the cavity (\unit[$\kappa/2\pi = 11$]{GHz}), effectively reducing the Purcell factor and ii) a decreased average resonant driving efficiency, leading to a vanishing SLR and precluding RF measurements. In the case presented in Fig.\,\ref{fig3}c, the fit yields a SLR of 14.0, corresponding to a contrast of 93.3\%. To our knowledge, it constitutes the only report so far of RF measured with a SLR exceeding 1 without relying on temporal filtering~\cite{Hensen2015,Tran2017,Ruf2021}.

RF as a function of cavity displacement is shown in Fig.\,\ref{fig3}d. The top panel represents the condition $\Delta_\mathrm{NV}=0$. For $\Delta_\mathrm{cav}=0$, both the laser and mode M2 are on-resonance with \Ex. For moderate cavity displacement, the driving field is approximately constant since the laser only interacts with the cavity via mode M1. Thus, for small values of $\Delta_\mathrm{cav}$, only the coupling strength between \Ex\ and mode M2 varies, leading to the observation of a strong RF peak. As the cavity displacement is further increased in the negative direction, a second peak emerges, corresponding to the laser being resonantly enhanced by mode M1. The drive enhancement increases the excited state population, but simultaneously promotes non-cycling mechanisms such as spin-flips and ionization. Concomitantly, the coupling between \Ex\ and M2, the collection mode, is reduced, resulting in a drop in RF signal and a decrease of the Purcell enhancement. A combination of these effects explains the difference in prominence between the peaks associated with modes M1 and M2: for a large-enough laser power, the signal gain obtained due to a stronger drive does not compensate the reduction induced by the cavity displacement.

The bottom panel of Fig.\,\ref{fig3}d shows the effect of introducing detuning on the same experiment ($\Delta_\mathrm{NV}=\mathrm{\unit[2\pi\cdot171]{MHz}}$, induced by the charge trap and thus measured in the same experimental run). The detuning effectively results in a reduction of the excited state population for otherwise identical parameters. Since the laser power is unchanged, the SLR decreases. However, the difference between the M1 and M2 peak heights vanishes, indicating that for this effective driving power, the response is linear: the loss of signal due to the cavity displacement is compensated by an increase in driving strength. The data shown in Fig.\,\ref{fig3}d are well-reproduced by taking into account two modes coupled with different $g_\mathrm{ZPL}$ values to one saturable transition subjected to charge noise~\cite{supp}. Using the experimentally-determined values for $\gamma_0$, $\Gamma_\mathrm{ext}$, $F_{\mathrm{P},\mathrm{M1}}$, $F_{\mathrm{P},\mathrm{M2}}$ and the laser background, we reduce the free parameters to two scaling factors, one for the collected signal and one for the input power.


\begin{figure*}
    \includegraphics[width=1\textwidth]{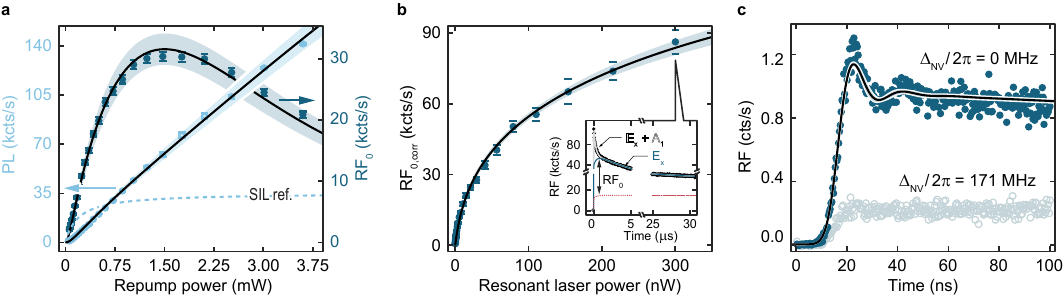}
    \caption{Cavity-enhanced zero-phonon emission. \textbf{a} ZPL count rate at the end of the repump pulse (empty light blue circles, left ordinate) and RF count rate (full teal circles, right ordinate) as a function of repump power. The data sets are jointly fitted with a rate-equation model (solid black lines). An example of state-of-the-art ZPL count rates for an \NV\ in a SIL, extracted from Ref.\,\citenum{Hensen2015}, is shown as a light blue, dashed line (left ordinate). \textbf{b} RF count rates as a function of resonant laser power. The rates correspond to the nominal amplitude of the \Ex\ signal and are corrected for laser background and decreasing repump efficiency. The data are fitted with a steady-state two-level saturation model including extrinsic broadening (black line). Inset: Slow RF decay for a laser power of \unit[300]{nW}. The total signal (solid grey line) includes \Ex\ fluorescence (teal line) as well as near-resonant fluorescence from the \A\, state and laser background (red dashed line). \textbf{c} RF at small time delays (\unit[300]{nW}, \unit[500]{ps} bin-time). For $\Delta_\mathrm{NV} = 0$ (full teal circles), the signal oscillations are well-captured by a model based on a driven two-level system (black line). A control experiment (blue-grey open circles) with $\Delta_\mathrm{NV}/2\pi = \unit[171]{MHz}$ shows no oscillations (washed out by charge noise and rise-time). On all graphs, shaded intervals (error bars) correspond to a 95\% confidence interval (1.96 standard deviations).}
    \label{fig4}
\end{figure*}

We turn our attention toward the evolution of PL and RF as a function of the repump laser power. The \NV\ is excited using the same sequence as depicted in Fig.\,\ref{fig3}b, with the laser, cavity and \Ex\ transition on resonance ($\Delta_\mathrm{cav}=\Delta_\mathrm{NV}=0$). The resonant ``probe'' pulse is kept at a constant power of \unit[5]{nW}. The PL signal (collected during the repump pulses) is shown in Fig.\,\ref{fig4}a (empty light blue circles, left ordinate). The increase in PL deviates in a subtle way from a linear behavior at high power. The maximal count rate obtained amounts to \unit[140.0]{kcts/s}, limited by the power density available to excite the NV. We stress that the photons collected here are exclusively ZPL photons, and reference our results against the ZPL intensities obtained for state-of-the-art measurements using an \NV\ embedded in a SIL, coated optics, and active aberration correction (dashed light blue line, left ordinate, reproduced from Ref.\,\citenum{Hensen2015}). The still unsaturated count rate measured with the cavity already exceeds the SIL performance by a factor four. 

The effect of increased repump power can also be seen in the RF signal (Fig.\,\ref{fig4}a, full teal circles, right ordinate). Since the resonant power is kept constant, the early RF signal ${RF_0}$ (before spin-shelving or ionization takes place, i.e. for times less than \unit[$\sim1$]{\textmu s}) is directly proportional to the shot-averaged joint probability of the NV to be in the correct charge (\NV) and spin (\msz) state. The linear increase of the $RF_0$ for repump powers below \unit[1.5]{mW} thus indicates an improvement in the \NV\ initialization. Combining this observation with the roughly linear increase in PL, whose intensity depends mostly on the product of the repump power and the \NV\ population, hints at a quicker recovery of the maximal charge population compared to its spin counterpart. The decrease in $RF_0$ beyond \unit[1.5]{mW} results from an optically-induced drift of \Ex\ in frequency, leading to $\Delta_\mathrm{NV}\neq 0$ and a consequentially reduced signal (see Fig.\,S3e,f in \citenum{supp}).

The PL and $RF_0$ signals are jointly fitted (solid black curves) using two coupled models: a 10-state model for the NV optical cycle under non-resonant excitation~\cite{Happacher2022}, and an effective 3-state model describing the population evolution during the resonant pulse (see Fig.\,S3 and \cite{supp} for details). The joint fit allows for a more robust retrieval of estimates for the PL saturated count rate $I^{PL}_\mathrm{sat}$ and saturation power $P^{PL}_\mathrm{sat}$. The regression yields \unit[$P^{PL}_\mathrm{sat}=67(37)$]{mW}, far from the maximal power experimentally available (\unit[3.6]{mW}), a clear evidence for the low power \emph{density} in our experiment. The estimated saturated count rate $I^{PL}_\mathrm{sat}$ is \unit[2.3(6)]{Mcts/s}, comparable to the best full-spectrum fluorescence rates using common non-resonant photonic structures~\cite{Hadden2010,Hensen2015, Hedrich2020}. The estimate is remarkably consistent with a cavity-modified $\xi_\mathrm{cav, \mathrm{M2}}= 0.45$ and $F_{\mathrm{P},\mathrm{M2}} = 1.8$: the cavity bandwidth restricts us to ZPL photons, but the cavity-enhanced branching ratio and emission acceleration compensate for the undetected PSB.

In Fig.\,\ref{fig4}b, we increase the resonant laser power and extract $RF_{0,corr}$, the \Ex\ transition RF intensity before shelving. In order to prevent power-induced drifts and potential degradation of the sample, the repump power is kept at \unit[0.2]{mW}.  Due to a combination of strain, charge noise and incomplete initialization, the close-by \A\, transition is also weakly driven for the highest resonant powers. The \Ex\ fraction of the signal is recovered from the total signal by fitting its bi-exponential decay: when excited to the \A\, state, the \NV\ is prone to undergo inter-system crossing (ISC), leading to a characteristic fast relaxation~\cite{Goldman2015a, Goldman2015b} (see inset). Further, for each measurement point, the residual laser background is subtracted and the $RF_{0,corr}$ signal is rescaled by the ratio between the measured PL signal over the PL signal at lowest resonant power. The latter step partially corrects for incomplete spin and charge initialization. For details on laser background determination and PL normalization, see \cite{supp}.

For the highest resonant power available, the $RF_{0,corr}$ intensity reaches \unit[86]{kcts/s}. The saturation behavior is modeled by convolving the steady-state response of a resonantly driven two-level system with the previously measured Lorentzian broadening:
\begin{equation}\label{equsat}
    RF_{0,corr} = I^{RF}_\mathrm{sat} \frac{2 a P^{RF}}{2 a P^{RF}+F_{\mathrm{P},\mathrm{M2}}^2\gamma_0^2+2\Gamma_\mathrm{ext}\sqrt{2 a P^{RF}+\gamma_0^2F_{\mathrm{P},\mathrm{M2}}^2}}.
\end{equation}
Here, the quickly decaying coherent response and the shelving processes have been disregarded, owing to the coarse time-binning and slow shelving. A scaling constant for the power (\unit[$a=1.7(5)\cdot10^3$]{MHz$^2$/nW}, accounting for all optical losses and coupling efficiencies) and the saturated intensity ($I^{RF}_\mathrm{sat}$) are left as the only free parameters. $I^{RF}_\mathrm{sat}$ amounts to \unit[250(30)]{kcts/s}, underlining once more the ability of the coupled \NV-cavity system to outmatch traditional photonic structures. The associated saturation power \unit[$P_\mathrm{sat}^{RF}=1.2(1)$]{\textmu W} (corresponding to half the saturated intensity) is predictably large as it would correspond to a Rabi frequency sufficient to overcome the charge noise-induced detuning distribution.

Finally, in Fig.\,\ref{fig4}c, we examine the time evolution of the (raw) RF signal while driving the \NV\ with the highest resonant power available (\unit[300]{nW}). For $\Delta_\mathrm{NV}=0$, coherent effects in the form of Rabi oscillations can be resolved. The oscillation contrast is reduced by residual laser background and \A\, transition fluorescence ($\sim$60\% of total signal), while charge noise and a slow pulse rise-time (\unit[9.5]{ns}, see \cite{supp}) contribute to a smearing of the oscillatory response. Taking into account these factors and considering the cycling character of the \Ex\ transition, we reproduce the data with a model based on the optical Bloch equations. The result of the fit, shown as a solid black line, yields the zero-detuning Rabi frequency \unit[$\Omega_{R,0}/2\pi = 51.1(26)$]{MHz}. A control experiment is done for a drive detuning of \unit[$\Delta_{NV}/2\pi = 171$]{MHz}. The fast oscillations for a detuned Rabi drive $\Omega_{R,171}$ are expected to be filtered out by the slow rise-time and charge noise, a prediction confirmed by our data: the smooth response rules out any artifacts stemming from the driving pulse shape in the zero-detuning dataset. The results in Fig.\,\ref{fig4}c demonstrate that the cavity-coupled \Ex transition can be driven coherently, opening opportunities for near-unity excitation probability using fast $\pi$ pulses.

\section*{Discussion} 

Comparing the extracted Rabi frequency on resonance $\Omega_{R,0} \sim 2.2\cdot \gamma_0F_{\mathrm{P},\mathrm{M2}}$ to the expected (based on the data fitted to Eq.\,\ref{equsat}) Rabi frequency at \unit[300]{nW}, $\Omega^{exp}_{R,0} \sim 5\cdot \gamma_0F_{\mathrm{P},\mathrm{M2}}$ indicates that the non-linear behaviour observed in Fig.\,\ref{fig4}b is not entirely due to two-level saturation. The premature saturation of \Ex\ is instead likely to be caused by a reduction in initial \msz\ population as the resonant power is increased (see Fig.\,S2, \citenum{supp}), explaining the difference between $I^{RF}_\mathrm{sat}$ and $I^{PL}_\mathrm{sat}$. While increasing the repump power improves the \NV\ initialization, non-resonant spin-pumping might be less efficient with the cavity on-resonance with the \NV\ upper orbital branch: the only state for which the spin-conserving decay probability increases significantly is \A, with spin-projection \mso\ implying a slower polarization rate into \msz. A more favorable path forward is enabled by the observation of the \A\, transition (see \cite{supp}) and would exploit resonant spin-pumping to achieve high-fidelity initialization~\cite{Robledo2011a}. Spin population initialization and control would limit the use of the green repump pulse to charge-state initialization. Implemented as previously demonstrated ``charge-resonance (CR) checks'' techniques~\cite{Hermans2023}, it would enable charge-noise-free operation. This would in turn lead to the generation of lifetime-limited photons and an automatic increase of the SLR by a factor $\sim \frac{\Gamma_\mathrm{ext}}{\gamma_0 F_\mathrm{P}}=6.9$, bringing it from our measured value of 14.0 (from Fig.\,\ref{fig3}) to $\sim$100, closer to the value needed to perform high-fidelity readout or spin-photon entanglement. We note that a reduced mode splitting would also have an important impact on the SLR, as the ratio of intra-cavity over propagating power would increase.

The results presented in this study were obtained at a location where diamond-induced losses limit the finesse to 4\,330 (for a bare-cavity finesse of $\mathcal{F}_{bare}=8\,880$), impacting the Purcell factor, $\beta$-factor, and ultimately the excitation-to-photon cavity outcoupling efficiency
\begin{equation*}
    \eta = \beta \frac{\kappa_{top}}{\kappa+\gamma_0}.
\end{equation*}
Here, $\kappa_{top}$ is the decay rate out of the top mirror. For our measured parameters, $\eta_{\mathrm{M2}}= 14.8\%$. The outcoupling efficiency could still be increased, considering the highest finesse  $\mathcal{F}=6\,700$ recorded on the same membrane (corresponding to a loss rate comparable to Ref.\,\citenum{HoyJensen2020}, albeit for an air-like mode). Operation at such a finesse would bring the Purcell factor to $F_{P} = 2.2$ and the efficiency to $\eta=28\%$. The limit to our collection efficiency ($\sim$10\%), while set in part by the silicon APDs ($\sim$70\%), stems mostly from the optics in the path, in particular by the objective transmission ($\sim$45\%) and aberrations. More mature implementations with quantum dots present an optimistic perspective~\cite{Tomm2021,Ding2023}, bringing the realization of a total (end-to-end) efficiency $\Sigma\sim 15\%$ within reach, and consequently a tremendous improvement over current \NV-based spin-photon interfaces.

In conclusion, we have demonstrated an efficient photonic interface to an isolated, narrow-linewidth \NV\ center, a spin-based qubit in diamond with a proven track-record. With Purcell factors of up to 1.79, the interaction of the \NV\ with light is profoundly altered and the coherent emission fraction is increased by more than an order of magnitude, from 3\% to 44.7\%. This allows us to measure ZPL count rates exceeding \unit[140]{kcts/s} under non-resonant excitation and resonance fluorescence count rates reaching \unit[86]{kcts/s} with a signal-to-laser ratio largely exceeding unity, without any temporal filtering. Our work represents crucial progress in addressing a problem which has until now stymied the development of prototypical quantum networks and limited their extension beyond a few nodes~\cite{Pompili2021,Hermans2022}. Projecting the current system efficiency to spin-spin entanglement rates, our system could already increase the success probability of one (two) photon protocols by more than one (two) order of magnitudes~\cite{Barrett2005,Hensen2015,Pompili2021,Hermans2022}, opening opportunities for the implementation of the \NV-cavity platform as an efficient building block for quantum networks~\cite{Childress2006,Rozpedek2018,Wehner2018}.

\section*{Data availability}
The data that support the findings of this study are available from the
corresponding authors on reasonable request.

\section*{Methods}
\subsection*{Cavity and sample}
The top mirror curvature is created via CO$_2$ laser ablation following a previously developed method~\cite{Greuter2014,Najer2017}. Aberrations in the ablating laser beam leads to a slight ellipticity (typically $\sim$5\%) of the crater. The top (bottom) DBR mirror transmission amounts to \unit[$\mathcal{T}_{top}=485$]{ppm} (\unit[$\mathcal{T}_{bottom}=56$]{ppm}). Both mirrors are high-index terminated, with a stopband centered around \unit[637]{nm} and $<30\%$ reflective at \unit[532]{nm}. The additional scattering and absorption losses at the measurement site amount to \unit[$\mathcal{L}=908.7$]{ppm}.

NV centers are created in \unit[20-by-20]{\textmu m$^2$} membranes by implantation post-fabrication as described in Ref.\,\citenum{Yurgens2022}. The ions ($^{12}\mathrm{C}^+$) are implanted with an energy of \unit[50]{keV} and a fluence of \unit[$5\cdot10^8$]{cm$^{-2}$}, aiming at forming sparse NVs at a depth of \unit[$\sim$66]{nm}, close to a field antinode in a subsequently assembled cavity. Membranes are released by micromanipulators onto a bottom mirror where they are bonded via van der Waals interactions to the mirror's surface (implanted side against the mirror). The bottom mirror is mounted on a set of XYZ piezosteppers (attocube ANP51). The top mirror is bolted onto a surrounding frame and adjusted in order to achieve mirror parallelity \unit[$<$0.4]{mrad}~\cite{Barbour2011}. The full cavity is supported by another set of XYZ steppers (attocube ANP101), placed in a housing under a fixed three-lens objective and suspended in a tube filled with \unit[20-30]{mbar} of high-purity helium. The entire tube is cooled in a super-insulated helium-bath cryostat to \unit[$\sim 4$]{K}.

\subsection*{Setup}
The sample is illuminated non-resonantly either by a low-coherence laser at \unit[532]{nm} (CW and RF experiments) or the \unit[5]{nm}-filtered output of a pulsed (\unit[$\sim$70]{ps}) supercontinuum laser (NKT Photonics SuperK, lifetime experiments). Resonant excitation is done with a CW narrow-linewidth external-cavity diode laser (Toptica DLPro) locked to a wavemeter. Both CW lasers are intensity-gated by AOMs (Gooch \& Housego) in a double-pass configuration, providing an isolation of \unit[$>60$]{dB}. The collected fluorescence is long-pass-filtered (\unit[$\sim 594$]{nm}), optionally bandpass-filtered ($\sim$\unit[$636 \pm 4$]{nm}, Semrock), coupled to a fiber, and detected using either one APD, a pair of APDs (Excelitas SPCM-AQRH), or a diffraction grating spectrometer and liquid nitrogen-cooled back-illuminated CCD (Princeton Instruments Acton 2500i and Pylon camera). The signal from the APDs is recorded either in histogram or time-tagged mode by a timing module (Picoquant Picoharp 300). The triggers for the timing module and AOM pulsing are sent by a dual-output function generator (Agilent 33500B). All DC voltages are generated by a 24-bit digital-to-analog card (Basel Precision Instruments), amplified if needed by low-noise 10x amplifiers (Electronics workshop, University of Basel).

\subsection*{PL and RF measurements}
The spectra in the scans presented in Fig.\,\ref{fig1} are recorded with integration times \unit[1-3]{s} at a non-resonant power of \unit[$\sim$1]{mW}. The high-resolution scan is dispersed using a high-density (\unit[2\,160]{grooves/mm}) grating, resulting in a slightly asymmetric peak. The autocorrelation signal is recorded using a Hanbury Brown-Twiss setup for a duration of \unit[20]{min} at a non-resonant power of \unit[$\sim$1.9]{mW}. An automatic ``relock'' protocol corrects for drifts in $\Delta_\mathrm{cav}$ every \unit[5]{min}. Each decay trace as a function of $\Delta_\mathrm{cav}$ in Fig.\,\ref{fig2} is integrated for \unit[90]{s}, with a cavity ``relock'' every \unit[5]{min}. The RF scans in Fig.\,\ref{fig3}c and d are averaged from sets of 50 and 70 scans, respectively, recorded with \unit[0.1]{s} integration time per point. During each scan, the nearby trap can become loaded or unloaded. Histogramming the counts for each ($\Delta_\mathrm{V}$ or $\Delta_\mathrm{cav}$) setpoint yields a bi-modal distribution, allowing the attribution of each of the scan setpoints to either the ``loaded'' or ``unloaded'' state. The data sets used in Fig.\,\ref{fig4} are acquired by time-tagging the start of each non-resonant pulse on one channel of the timing module and the arrival of all photons on another channel. The integration time for each power setpoint is \unit[10]{min}, after which a ``relock'' sequence is run. The time-tagged data can then be recast as average intensities per pulse (allowing discrimination of the trap state), or as averaged time-traces.

\begin{acknowledgments}
The authors thanks Lilian Childress for fruitful discussions. The authors acknowledge financial support from the National Centre of Competence in Research (NCCR) Quantum Science and Technology (QSIT), a competence center funded by the Swiss National Science Foundation (SNF), from Swiss SNF Project grant No.\,188521, from the Swiss Nanoscience Institute (SNI), and from the Quantum Science and Technologies at the European Campus (QUSTEC) project of the European Union’s Horizon2020 research and innovation program under the Marie Sklodowska-Curie Grant Agreement No.\,847471.
\end{acknowledgments}

\bibliography{biblio_short} 
\end{document}


\newcommand{\NV}{\ensuremath{\text{NV}^{\mathrm{-}}}\ }
\newcommand{\NVo}{\ensuremath{\text{NV}^{\mathrm{0}}}\ }
\newcommand{\Ex}{\ensuremath{\text{E}_{\mathrm{x}}}\ }
\newcommand{\Ey}{\ensuremath{\text{E}_{\mathrm{y}}}\ }
\newcommand{\A}{\ensuremath{\text{A}_{\mathrm{1}}}\ }
\newcommand{\msz}{\ensuremath{\text{m}_{\mathrm{s}}=0}\ }
\newcommand{\mso}{\ensuremath{\text{m}_{\mathrm{s}}=\pm 1\ }}

\renewcommand{\figurename}{Fig.}
\renewcommand{\thefigure}{S\arabic{figure}}

\renewcommand{\theequation}{S\arabic{equation}}
\renewcommand{\bibnumfmt}[1]{[S#1]}
\renewcommand{\citenumfont}[1]{S#1}

\preprint{AIP/123-QED}


\title[Cavity-assisted resonance fluorescence from a nitrogen-vacancy center in diamond]{\textbf{Supplementary information: Cavity-assisted resonance fluorescence from a nitrogen-vacancy center in diamond}} 
\author{Viktoria Yurgens}
\thanks{These two authors contributed equally to this work}

\author{Yannik Fontana}
\email[]{yanniklaurent.fontana@unibas.ch}
\thanks{These two authors contributed equally to this work}

\author{Andrea Corazza}

\author{Brendan J. Shields}
\altaffiliation[Present address: ]{Quantum Network Technologies,  Boston, Massachusetts 02215, USA}

\author{Patrick Maletinsky}

\author{Richard J. Warburton}
\affiliation{Department of Physics, University of Basel, CH-4056 Basel, Switzerland}

\date{\today}

\maketitle

\section{Cavity characteristics}
The cavity described in this manuscript is close to one-sided, with almost 90\% of the unscattered or absorbed intra-cavity power leaking through the top mirror ($\mathcal{T}_\mathrm{top}=\unit[485]{ppm}$ vs $\mathcal{T}_\mathrm{bot}=\unit[56]{ppm}$). All measurements are realized on a single \NV center, located at a position where the cavity finesse is 4\,330 (a reduced value compared to the finesse of 8\,880 for the bare cavity, which itself is slightly lower than the cavity design finesse of 11\,614). The longitudinal mode to which the NV couples is always the same, and corresponds to the nominally degenerate $m_{(0,0,16)}$ mode. As seen in Fig.\,1b, other modes corresponding to a shorter cavity (positive $\Delta_\mathrm{cav}$) are accessible. Using these modes would theoretically lead to slightly larger $\mathrm{NV^-}$\!-cavity coupling strength. However, mode $m_{(0,0,14)}$ corresponds to the contact mode (top mirror touching the bottom mirror), and mode $m_{(0,0,15)}$ was therefore left as a buffer mode to avoid even slight contact between the mirrors.

The mode-splitting of $m_{(0,0,16)}$ into modes M1 and M2 is ever-present due to a slightly non-spherical ($\sim 5\%$ ellipticity) top mirror. On the bare cavity, this mode-splitting is typically not resolved in a scan of the probe laser frequency or cavity length, as our top mirror anisotropy would result in a doublet splitting of $<\unit[1]{GHz}\ll \kappa = \unit[11]{GHz}$. On the diamond, on the other hand, it varies depending on the exact location, indicating that the dominant mode splitting mechanism originates from birefringence in the diamond or, most likely, caused by the diamond).

The cavity resonances are characterized by sweeping the cavity through a main fixed laser carrier and a sideband, the sideband providing the required reference in frequency to convert displacements into frequencies. The cross-polarization arrangement of our optics, formed by a quarter- and half-waveplate (QWP, HWP), allows us to convert a change in phase due to coupling to the cavity into a change in intensity in the collection path. Setting the QWP so that the light impinging on the cavity is circularly polarized maps to a balanced-cavity (a cavity with equal-reflectivity mirrors) reflection measurement, while aligning one QWP principal axis with the incoming beam leads to two main scenarios:
\begin{enumerate}
    \item The HWP rotates the polarization of the laser so that it bisects the angles set by the polarization axis of modes M1 and M2. In this case, the field projection on modes M1 and M2 will acquire different phases for the same $\Delta_\mathrm{cav}$, leading to a rotation of the reflected light's polarization and thus optical contrast. Since no signal can pass through the crossed polarizers if both modes are detuned with respect to the laser, this configuration maps to a balanced cavity transmission measurement.
    \item The HWP rotates the polarization of the laser so that it coincides with the polarization axis of mode M1 or M2, leading to a vanishing rotation of the reflected field and its filtering by the cross-polarized arrangement.
\end{enumerate}
The second situation is forms the basis of the laser suppression scheme used in this work.

\section{Second-order autocorrelation}
The model used to fit the $g^{(2)}(\tau)$ data is based on an incoherently-driven rate equation model~\cite{Berthel2015}. The continuous green excitation justifies in large part the restriction to three effective levels, as it continuously pumps the \NV spin in \msz and most of the remaining rates can be lumped into effective rates.
Defining the three state populations (ground, excited and shelving state) as $\rho=[\rho_g, \rho_e, \rho_s]^T$, the matrix governing the system dynamics is:
\begin{equation}
\dot{\rho}(t)=
\begin{bmatrix}
    -k_{532} & k_{eg} & k_d \\
    k_{532} & -(k_{eg}+k_s) & 0 \\
    0 & k_s & -k_d \\
\end{bmatrix}
\cdot \rho(t).
\end{equation}
The system can be solved analytically or numerically. The way the autocorrelation is measured (time-difference between consecutive photons) gives the initial condition at time $\tau=0$, $[\rho_g(0),\rho_e(0),\rho_s(0)]^T=[1,0,0]^T$. The steady-state population $\rho_e(\infty)$ can be used to normalize $\rho_e(\tau)$, yielding $g^{(2)}(\tau) = \frac{\rho_e(\tau)}{\rho_e(\infty)}$.

The result can be further simplified considering that $\{k_s,k_d\}\ll k_{532}+k_{eg}$ so that
\begin{widetext}
\begin{equation}
  g^{(2)}(\tau) = 1-(1+\frac{k_{532}k_s}{k_d(k_{eg} + k_{532})})e^{-(k_{eg}+k_{532})t} + \frac{k_{532}k_s}{k_d(k_{eg} + k_{532})} e^{-(k_d + \frac{k_{532}k_s}{k_{532}+k_{eg}})}.
\end{equation}
\end{widetext}
Note that the contribution from a classical photoluminescence background $\mathcal{B}$ can be also be added using $g^{(2)}_\mathcal{B}(\tau) = \frac{\rho_e(\tau)+\mathcal{B}}{\rho_e(\infty)+\mathcal{B}}$.

\section{Purcell factor}\label{Purcellfactor}
Two main factors hinder a straightforward analysis of the lifetime reductions observed while varying $\Delta_\mathrm{cav}$: the possibility for the \NV transition to couple to two non-degenerate modes M1 and M2, and the presence of vibrations inducing involuntary excursion in displacement $\delta$. Considering the vibration-induced excursions as normally distributed (supported by calibration measurements) such that the probability distribution reads:
\begin{equation}
     P_\mathrm{vib}(\delta) = \frac{1}{\sqrt{2\pi}\sigma_\mathrm{vib}}e^{-\frac{1}{2}\bigl(\frac{\delta}{\sigma_\mathrm{vib}}\bigr)^2}
\end{equation}
and the fact that modes M1 and M2 are orthogonal to each other, we can write the emission rate $R(\tau)$ during the period following a fast excitation pulse as:
\begin{widetext}
\begin{equation}
  R(\tau) \propto
  \int\limits_{0}^{\infty} P_\mathrm{vib}(\delta) \cdot 
  \rho_{e} \cdot \gamma_0 \cdot\Bigl(\frac{\sin^2(\theta_\mathrm{cav})}{1+\bigl(\frac{\Delta_\mathrm{cav}+\Delta_\mathrm{M1}+\delta}{W_\mathrm{cav}}\bigr)^2} + \frac{\cos^2(\theta_\mathrm{cav})}{1+\bigl(\frac{\Delta_\mathrm{cav}+\delta}{W_\mathrm{cav}}\bigr)^2}\Bigr) \cdot C \cdot
  e^{-\gamma_0 \Bigl[\Bigl(\frac{\sin^2(\theta_\mathrm{cav})}{1+\bigl(\frac{\Delta_\mathrm{cav}+\Delta_\mathrm{M1}+\delta}{W_\mathrm{cav}}\bigr)^2} + \frac{\cos^2(\theta_\mathrm{cav})}{1+\bigl(\frac{\Delta_\mathrm{cav}+\delta}{W_\mathrm{cav}}\bigr)^2}\Bigr) C +1\Bigr] \tau} \mathrm{d}\delta
  \label{decayall}
\end{equation}
\end{widetext}
with $\rho_e$ the excited state probability, $\gamma_0$ the bulk-like, natural decay rate of the $\mathrm{NV^-}$, $\theta_\mathrm{cav}$ the angle with respect to mode M2 of the projection of the transition dipole on the plane defined by the electric fields of modes M1 and M2, $\Delta_\mathrm{cav}$ and $\Delta_\mathrm{M1}$ the detuning in displacement with respect to mode M2 and the splitting between M2 and M1, respectively, $W_\mathrm{cav}$ the HWHM of the cavity modes and $C=F_\mathrm{P,degen}-1$ where $F_\mathrm{P,degen}$ would correspond to the total Purcell factor (as defined in the main text) if both modes were degenerate. The description given above implicitly eliminates the cavity decay, justified as even for large Purcell factors $\tau_\mathrm{cav}^{-1}\gg \gamma_0 F_\mathrm{P,degen}$.

Finally, the finite rise-time of the laser pulse and detector response, the instrument response time $\mathcal{J}(\tau)$, can be included by convolution:
\begin{equation}
    \tilde{R}(\tau) = R(\tau)\circledast \mathcal{J}(\tau).
\end{equation}
While Eq.\,\eqref{decayall} gives the total decay rate, it can readily be recast into $R_\mathrm{M1}(\tau)$ and $R_\mathrm{M2}(\tau)$, the contributions from each modes separately. For an arbitrary alignment of the detection channel polarization with respect to mode M2, $\theta_\mathrm{det}$, and considering a scaling factor $\zeta$ for the detection efficiency, the intensity at the detector as a function of time is:
\begin{equation}
    I_\mathrm{det}(\tau)= \zeta \cdot \bigl(\sin^2(\theta_\mathrm{det})\tilde{R}_\mathrm{M1} + \cos^2(\theta_\mathrm{det})\tilde{R}_\mathrm{M2}\bigr).
\end{equation}
We note that the QWP is set so that its action is the identity, and thus plays no role. This result can then be numerically integrated to fit the full lifetime versus cavity displacements dataset and obtain the results shown in Fig.\,2 of the main manuscript (light blue curves).


\begin{figure*}
    \includegraphics[width=1\textwidth]{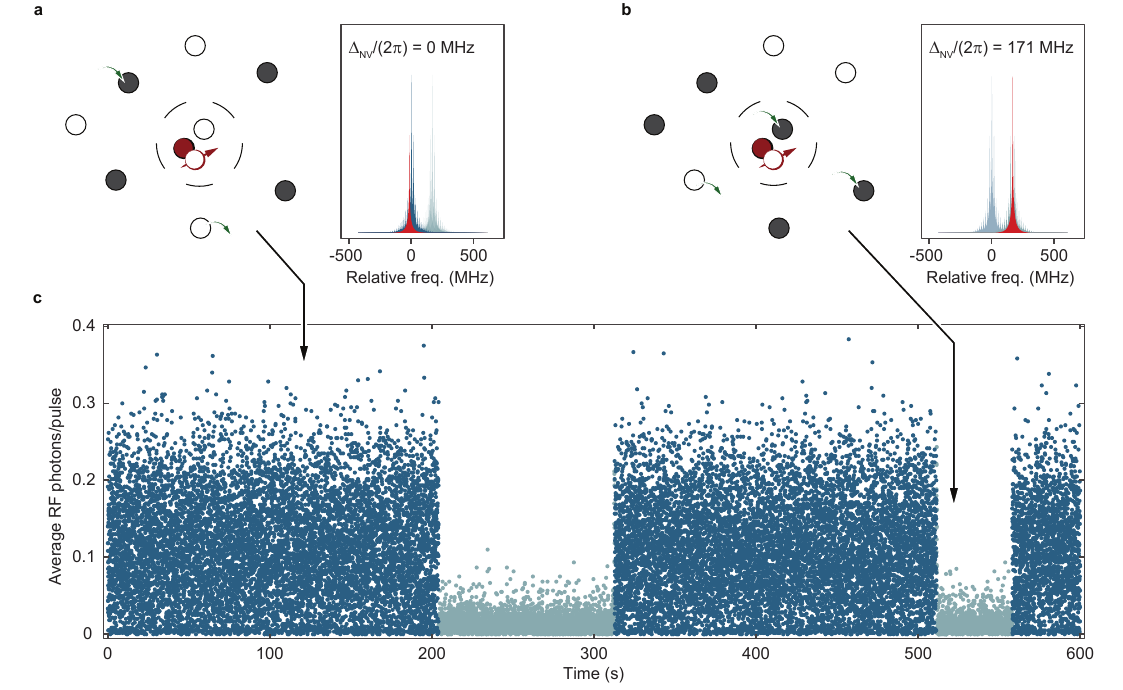}
    \caption{\textbf{Resolved Stark shift.} \textbf{a} Left: A \NV center (red and white) surrounded by charge traps, either loaded (grey) or unloaded (white). Each repump pulse has a probability to produce loading or unloading events (green arrows). One of the charge traps is particularly close to the $\mathrm{NV^-}$. Right: A combination of the nearby trap state (unloaded) and the configuration of the remaining ensemble results in the \Ex transition (red) appearing within the lower frequency sub-distribution. \textbf{b} Left: Same as \textbf{a}, but with the nearby trap loaded. Right: the nearby trap shifts the \Ex transition globally, while the particular configuration of the ensemble results in a (unresolved) shift within the higher frequency sub-distribution. \textbf{c} Evolution of the RF over long times, with the laser on resonance with the lower frequency sub-distribution.
    }
    \label{Sfig0}
\end{figure*}

\section{Resonance fluorescence}
\subsection*{Single and ensemble charge noise}
As seen in the main text, the systematic application of a green repump pulse before any resonant laser pulse leads to the measurement of the extrinsically broadened linewidth, i.e.\ the linewidth broadened shot-to-shot by the stochastic change of charge state of traps in the diamond material. In our particular case, the effect of charge noise can be separated into two distinct categories sharing the same underlying physical mechanism: the ``unresolved'' effect of an ensemble of traps (the normal scenario for all emitters prone to charge noise broadening), and the ``resolved'' effect of a single trap (further referred as ``the trap'') particularly close to the $\mathrm{NV^-}$.
This situation is depicted in Fig.\,\ref{Sfig0}a,b. Fig.\,\ref{Sfig0}c shows a \unit[10]{min} time-trace made from averaging the signal per resonant pulse over 600 pulses (\unit[$\sim$10]{ms} averaging time) in order to increase the contrast. The state of the trap is color-coded as unloaded (teal) or loaded (blue-grey). The averaging minimally impacts the trap state determination due to infrequent jumps (\unit[>5]{mHz}). The ability to determine the trap state provides us with the opportunity to measure in single experiments signals with and without a fixed average detuning.

The unresolved part of charge noise is included in the models in two different computational ways:
\begin{enumerate}
    \item In Fig.\,3c of the main manuscript, it is phenomenologically decribed as creating a Lorentzian distribution in the detuning with respect to a fixed, centered laser. The deviation from a perfect Lorentzian distribution originates from residual laser background, itself having a Fano-like profile. We do not argue for a particular physical origin explaining the Lorentzian distribution and treat it mainly as a convenient heuristic to retrieve a FWHM \unit[$\Gamma_\mathrm{ext} / 2\pi=159.0$]{MHz}. The value determination is obtained by convoluting the Lorentzian noise distribution with the purely lifetime-limited linewidth of the Purcell-enhanced \Ex transition.
    
    \item For the other models, since a simple analytical form cannot be found, charge noise is introduced as Lorentzian-distributed detunings determined by $\Gamma_\mathrm{ext}$. The results including charge noise are then obtained by numerical integration over an interval $[-10\Gamma_\mathrm{ext}, 10\Gamma_\mathrm{ext}]$ or iteratively via sampling of the detunings until the averaging converges.
\end{enumerate}

\subsection*{E\textsubscript{x} and A\textsubscript{1} transitions}

\begin{figure*}
    \includegraphics[width=1\textwidth]{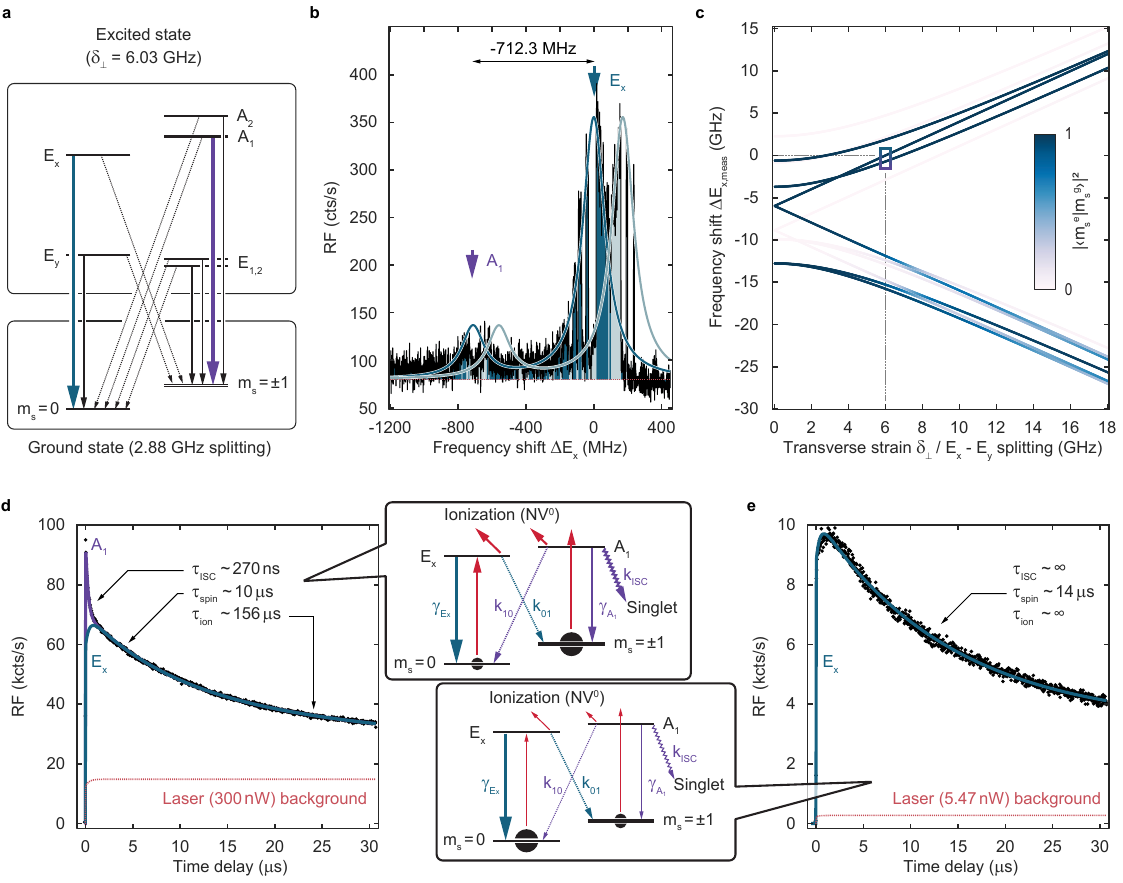}
    \caption{\textbf{Excited state and optical transitions.} \textbf{a} Optical transitions of an \NV subject to a transverse strain. Solid (dashed) arrows represent spin-conserving (spin-flipping) transitions. The \Ex (\A) transitions are displayed as thicker teal (purple) arrows. \textbf{b} RF spectrum of the \A and \Ex transitions. Both transitions stochastically hop between stable center frequencies depending on the the occupation of a nearby charge trap, giving the spectrum its particular jagged profile. \textbf{c} Calculated frequency shifts as a  function of transverse strain. The reference frequency corresponds to the \Ex $\rightarrow$ \mso  spin-conserving transition. The color scale corresponds to the transition strength. \textbf{d} RF signal decay for resonant driving of the \Ex transition at \unit[300]{nW}. The prominent, rapidly decaying peak is attributed to near-resonant driving of the \A transition which quickly undergoes ISC to the singlet (see diagram). \textbf{e} Same as \textbf{d} but at a much reduced (\unit[5]{nW}) power, precluding a strong driving of the \A transition and a direct signature of it at small time delays. The much weaker excitation rate also ensures that a significant fraction of the population remains within the \msz  manifold.
    }
    \label{Sfig1}
\end{figure*}

Until now, it was assumed that the transition examined in Fig.\,3 of the main manuscript corresponds to $\mathrm{m_s=0} \leftrightarrow \mathrm{E_x}$. We substantiate this claim hereafter. The \Ex transition is the one most likely to be observed in our condition (systematic repump) due to spin pumping in \msz and a high-degree of cyclicity even in the presence of transverse strain. The cyclicity itself is an outcome of the $\mathrm{E_x}$'s property to hybridize only marginally with other states in the upper (UB) or lower (LB) orbital branch of the excited manifold in the presence of transverse strain. This feature guarantees a protection against both inter-system crossing and spin-flip transitions to \mso ground states (both of which would render the transition unavailable for a period of time commensurate with their relaxation rate). However, the exact energies of any of the excited states do vary with transverse strain. In particular, transverse strain provokes the UB/LB splitting, and, together with the difference in eigenenergy of the \msz and \mso ground states, leads to the emergence of a variety of spin-conserving and spin-flipping transitions which can cross or anti-cross~\cite{Tamarat2008}. An example of all possible transitions for an NV subjected to transverse strain (leading to an \Ex/\Ey splitting $\delta_\perp$ of \unit[$\sim$6]{GHz}) is provided in Fig.\,\ref{Sfig1}. The solid (dashed) arrows represent spin-conserving (spin-flipping) transitions, with the \Ex transition highlighted in teal.

The strain Hamiltonian of the \NV is well known~\cite{Batalov2009,Maze2011,Doherty2011}, and so transverse strain and the observed transitions can be identified thanks to their particular splitting. The most common method relies on determining $\delta_\perp$ by the splitting between \Ex and $\mathrm{E_y}$, but in our case, a combination of strain and cavity-coupling precludes the observation of $\mathrm{E_y}$. Fortunately, other sets of transitions can be used, as shown in Fig.\,\ref{Sfig1}b where a small doublet of peaks can be spotted red-detuned from $\mathrm{E_x}$. The doublet nature of this and the \Ex peak is related to the nearby trap. Owing to the pre-characterization of this effect, we are able to fit the spectrum and retrieve the relevant splittings, in particular the splitting between the two transitions (\unit[$\Delta_\mathrm{E_x-A_1}=712.3$]{MHz}). This particularly small value is typical of the \A-\Ex transition frequencies crossing at \unit[$\delta_\perp \sim 5-7$]{GHz} (see Fig.\,\ref{Sfig1}c, showing all possible transition frequencies with respect to the observed \Ex frequency and color-coded by the spin-character overlap of their respective ground and excited states).

The small but measurable signal from \A can be explained by the incomplete initialization in \msz (as seen in the main text), providing a small population in $\mathrm{m_s}=\pm 1$, and the effect of the cavity, which helps to prevent spin and singlet shelving. The identification is further strengthened by observing the intensity decay of the RF signal at two different resonant laser powers (with the laser on resonance with \Ex). At high power (Fig.\,\ref{Sfig1}d), an initially fast decay can be observed, consistent with the documented characteristic of the \A state to quickly decay to the singlet by transverse spin-orbit-mediated intersystem crossing (ISC)~\cite{Maze2011,Goldman2015a,Goldman2015b}. Excitation of \A is assisted by a higher \mso population (due to incomplete recovery of spin-shelving), and near-resonant excitation (due to the small splitting) enhanced by power broadening and the effect of extrinsic broadening. At lower power (Fig.\,\ref{Sfig1}e), this fast component of the decay disappears: the \mso population is smaller (less efficient shelving for a constant off-resonant spin repumping rate) and the power-broadening is limited.

Finally, the characteristic decay time $\tau_\mathrm{spin}$ shown in Fig.\,\ref{Sfig1}d,e is consistent with the small probability of spin-flip of the \Ex transition for \NV centers with comparable strain~\cite{Robledo2011a}. The spin-shelving time is inversely proportional to the \Ex state population, a relation that we qualitatively observe in our dataset, but which is hard to quantify as the effect of charge noise distorts the observed distribution: until now, we considered $\Delta_\mathrm{NV}$, the laser detuning from \Ex averaged over many shots. However, the unresolved effect of the charge noise introduces small shot-to-shot detunings $\Delta_\mathrm{NV,shot}$. For $\Delta_\mathrm{NV,shot} \sim 0$, the probability of detecting a photon is higher, and the probability is high that such a photon will be detected early (spin-pumping is more efficient); for shots with $|\Delta_\mathrm{NV,shot}| > 0$, the total probability to detect a photon drops (lower excited state population), but the probability that this photon arrives late increases (slower spin-pumping).

\subsection*{Effect of cavity displacement}
The effect of cavity displacement while keeping the laser at a fixed frequency is shown in Fig.\,3d of the main text. The model for the fits hinges on the addressed transition coupling to both M1 and M2 with relative coupling strengths determined by $\theta_\mathrm{cav}$. The transition is modelled as a driven two-level system in the steady-state. While not formally accurate, this approximation holds due to a combination of the long probing time, as coherent effects fade after a few tens of nanoseconds at best (compared to a \unit[$\sim$31]{\textmu s} resonant pulse). Considering the laser-suppression configuration, the excitation laser can only occur through mode M1, while only the cavity population of mode M2 will reach the detector.

Denoting the normalized Lorentzian function for each mode as
\begin{equation}
    L_1(\Delta_\mathrm{cav}) = \Bigl(1+\bigl(\frac{\Delta_\mathrm{cav}+\Delta_\mathrm{M1}}{W_\mathrm{cav}}\bigr)^2\Bigr)^{-1} \quad \mathrm{and} \quad L_2(\Delta_\mathrm{cav}) = \Bigl(1+\bigl(\frac{\Delta_\mathrm{cav}}{W_\mathrm{cav}}\bigr)^2\Bigr)^{-1}.
\end{equation}
The occupation probability of the excited state (neglecting initial coherent effects) becomes
\begin{equation}
    \rho_e = \frac{1}{2} \cdot \frac{\frac{1}{2}P}{\frac{1}{2}P+\delta_\mathrm{cn}^2+\bigl(\frac{1}{2}\gamma_\mathrm{cav}\bigr)^2}=
    \frac{1}{2} \cdot \frac{\frac{1}{2}P_0L_1}{\frac{1}{2}P_0L_1+\delta_\mathrm{cn}^2+\frac{1}{4}\Bigl(\bigl(L_1C_1+L_2C_2+1\bigr)\gamma_\mathrm{cav}\Bigr)^2}
\end{equation}
with $P_0$ the maximal intra-cavity power on resonance, $\delta_\mathrm{cn}$ the detuning introduced by charge noise and $C=F_\mathrm{P}-1$ the effect of each mode on de-excitation, as previously described. We note that the effect of charge noise on the mode-coupling is neglected as $\Gamma_\mathrm{ext}\ll \kappa$.

Finally, the response of the system to cavity detuning can be written as
\begin{equation}
    I_\mathrm{det}\bigl(\Delta_\mathrm{cav},P_0\bigr) = \mathcal{A} \cdot L_2C_2\gamma_0 \Bigl(
    \int_{-\infty}^\infty \frac{\rho_e(\delta_\mathrm{cn},P_0,\Delta_\mathrm{cav})}{\pi\Gamma_\mathrm{ext}\bigl(1+(2\delta_\mathrm{cn}/\Gamma_\mathrm{ext})^2\bigr)}d\delta_\mathrm{cn}
    \Bigr)
\end{equation}
with $\mathcal{A}$ a scaling constant accounting for the total efficiency, and noting that $P_0 \propto \mathcal{F}$. The interplay of the excitation and detection explains the behaviour shown in Fig.\,3. At very low (effective) power or large frequency detuning, when the response of the transition stays linear and as the cavity mode M1 (M2) comes into resonance with the transition and laser, the increase (decrease) in excitation efficiency through mode M1 compensates a reduced (increased) coupling to the collection mode M2. On the other hand, if the effective power is large enough, the transition will saturate when the cavity is resonant with mode M1 while being still largely in the linear regime when the cavity is resonant with mode M2, producing an unbalanced doublet.

\section{Saturation analysis }
\subsection*{Non-resonant repump laser: rate equations model and drifts}
\begin{figure*}
    \includegraphics[width=1\textwidth]{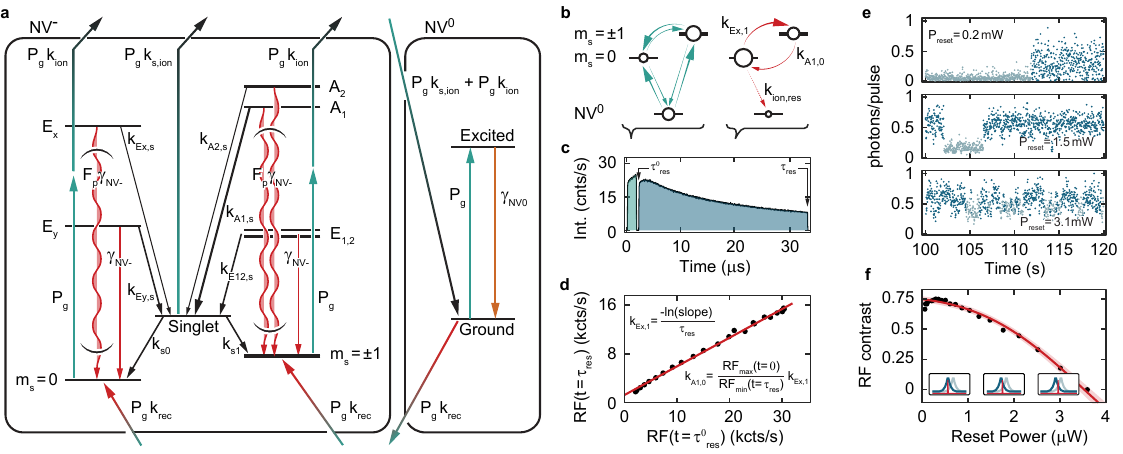}
    \caption{Optical cycle and model. \textbf{a} Full 10-state model used to simulate the repumping mechanism. Off-resonant excitation of all \NV excited states and the lumped NV$^\mathrm{0}$ excited state is represented by green arrows. Green-to-black (green-to-red) arrows represent the second step of the 2-photons process required to convert \NV to NV$^\mathrm{0}$ and vice-versa. The \NV singlet is represented by a single state, with black arrows from the \NV triplet excited states to the singlet representing state-dependent non-radiative ISC. The singlet can in turn decay back to the \NV triplet ground state, increasing the \msz and \mso spin sublevel population with equal rates. Direct, spin-conserving transitions from the \NV triplet excited states to the triplet ground state are represented by red arrows, and have nominally equal rates, modified for the upper orbital branch by the effect of the cavity via the Purcell effect. The NV$^\mathrm{0}$ electronic structure is simplified and represented by single ground and excited states, pumped non-resonantly and decaying radiatively (orange arrow) at a fixed rate. \textbf{b} Schematics of the effective population transfer within the triplet ground state during the repump pulse (left, green arrows) and the resonant laser pulse (right, red arrows). \textbf{c} Typical measurement showing the count rate increase due to de-ionization and polarization during the repump pulse (green filling) and subsequent decrease during the resonant laser pulse due to spin-shelving and ionization. Near-resonant excitation of the \A transition explains incomplete spin-shelving into $\mathrm{m_s}=\pm 1$. The laser powers for these measurements are \unit[605]{\textmu W} (repump laser) and \unit[5]{nW} (resonant laser). \textbf{d} RF signal corresponding to the tail of the resonant pulse as a function of RF signal at the onset of the resonant pulse (black circles). A linear regression (red line) allows the estimation of the \msz to \mso spin-shelving rate as well as the weak converse \mso to \msz rate. \textbf{e} Time-traces of the RF at different repump powers. The teal (light blue) dots correspond to the \NV being driven while the nearby charge trap is either loaded or unloaded, leading to no (teal) or a finite (light blue) detuning between the \Ex transition and the resonant laser. One striking feature is the decrease in contrast as the repump power is increased. \textbf{f} Decrease in RF contrast (black dots), explained by a repump-power dependent quasi-static field shifting the \NV transitions and inducing a progressive detuning of the resonant laser (see insets). The effect is well captured by a simple model (red line) considering a linear field change leading to a linear excited state shift provoked by a change in repump power.}
    \label{Sfig3}
\end{figure*}

In order to model the optical cycle of the NV center during the interleaved green and red pulses, we develop a model based on rate equations for both the off-resonant repump and resonant probe sub-cycles.

The important rates to consider during the repump pulses are displayed in Fig.\,\ref{Sfig3}a, leading to a 10-state model. The non-resonant character of the green pulses leads to a rather complicated picture where all excited states (with the exception of $\mathrm{E_1}$ and $\mathrm{E_2}$ lumped together as $\mathrm{E_{1,2}}$) can be populated and radiate. Additionally, the singlet state needs to be incuded, albeit in a simplified form (one state). The ionized \NVo state also plays a role and is included as a ground and excited state. Thankfully many of the rates needed for the model can be extracted from the literature or are measured independently. 

All rates, except for the recombination rate $k_\mathrm{rec}$ and green power scaling factor $\beta_{532}$, are fixed. While the spin-flipping transitions are ignored (low probability), inter-system crossing rates are needed and adjusted for our strain-induced level of mixing using reference values from Ref.\,\citenum{Goldman2015a}. An extra scaling factor for the green power is considered for transitions pertaining to the UB and LB of the excited manifold. We consider the following approximations: the UB dipoles (or major axis) are taken as contained in the plane defined by the M1 and M2 mode polarization (100), while the LB dipoles are primarily orthogonal to both the UB dipoles and the \NV quantization axis (along $[111]$ or equivalent). The angle and projection between each branch orientation and the green laser polarization can then be inferred considering our knowledge of $\theta_\mathrm{cav}$ and the effect of the optics on the green polarization state before entering the objective (elliptical and principally oriented almost perpendicular to the red laser due to different retardation of the \unit[633]{nm} optimized waveplates).

The bulk-like \NV recombination rate $\gamma_0$ (corresponding to $\gamma_\mathrm{NV^-}$ in Fig.\,\ref{Sfig3}a) was determined in the main text, together with the Purcell factor applied to the UB states (the LB states are far detuned from the cavity resonances). All other rates are taken from Ref.\,\citenum{Happacher2022}, including the newly proposed one-photon ionization process from the singlet.
All the rates can be cast into a single 10-by-10 matrix $M_G$ and the population of all states summed-up in a state-vector $\mathrm{p}$. The evolution of $\mathrm{p}$ from the beginning ($\tau_\mathrm{G}^0$) to the end ($\tau_\mathrm{G}$) of the repump pulse is then
\begin{equation}
    \mathrm{p}(\tau_\mathrm{G})=e^{M_G \cdot (\tau_\mathrm{G}-\tau_\mathrm{G}^0)}\mathrm{p}(\tau_\mathrm{G}^0).
\end{equation}

The situation is much simpler during the red resonant pulse, as the resonant power is constant. Assuming that most transitions are not excited, the dynamics can be modelled as two effective rates $k_\mathrm{E_x,1}$ and $k_{A_1,0}$ coupling directly \msz and \mso and working one against each other. Additionally, the possibility for ionization is included with the rate $k_\mathrm{ion,res}$ (see Fig.\,\ref{Sfig3}b). The variation in the non-resonant power allows us to vary the initial population of \msz and sample its relative decrease (see Fig.\,\ref{Sfig3}c), allowing the extraction of values for $k_\mathrm{E_x,1}$ and $k_\mathrm{A_1,0}$ (Fig.\,\ref{Sfig3}d) and leaving $k_\mathrm{ion,res}$ as the only undetermined parameter for the resonant part of the pulse sequence. As for the non-resonant case, an evolution matrix $M_\mathrm{res}$ for $\mathrm{p}$ can be built:
\begin{equation}
    \mathrm{p}(\tau_\mathrm{res})=e^{M_\mathrm{res} \cdot (\tau_\mathrm{res}-\tau_\mathrm{res}^0)}\mathrm{p}(\tau_\mathrm{res}^0).
\end{equation}
Since the excited states are bypassed by our effective coupling model, most of the elements of the matrix will be zero.

Finally, the waiting time between each pulse can be included in the same fashion, giving a matrix $M_\mathrm{wait}$ governing the de-excitation processes only (radiative and non-radiative). The evolution over a full sequence unit is then:
\begin{equation}
    \mathrm{p}(\tau_\mathrm{unit})=e^{M_\mathrm{wait} \cdot (\tau_\mathrm{wait}-\tau_\mathrm{wait}^0)}
    e^{M_\mathrm{res} \cdot (\tau_\mathrm{res}-\tau_\mathrm{res}^0)}
    e^{M_\mathrm{wait} \cdot (\tau_\mathrm{wait}-\tau_\mathrm{wait}^0)}
    e^{M_\mathrm{G} \cdot (\tau_\mathrm{G}-\tau_\mathrm{G}^0)}
    \mathrm{p}(\tau_\mathrm{unit}^0)
     = e^{M_\mathrm{unit}}\mathrm{p}(\tau_\mathrm{unit}^0).
\end{equation}

The evolution of the population (and thus the recorded PL and RF intensities) at various times along the pulse can then be evaluated considering that the system will reach a steady-state such that
\begin{equation}
    \dot{\mathrm{p}}(\tau_\mathrm{unit}) = 0 = \frac{d}{dt}\bigl(e^{M_\mathrm{unit}}\bigr)\mathrm{p}(\tau_\mathrm{unit}^0).
\end{equation}

The steady-state can be directly computed or the system can be initialized and iterated until convergence. We choose the latter option as only a few iterations are necessary and the numerical calculation are more stable. The entire procedure can be iterated over in order to generate the fits (for the remaining unknown rates and scaling factors) displayed in Fig.\,4a of the main text.

Finally, we note that one tweak to the model is needed: the increase in green power results in an uncompensated drift of the transitions. While the exact origin of this drift is unknown, we suspect an accumulation of charges at the interfaces of the diamond membrane, fueled by ionization of other defects in the sample. As seen in Fig.\,\ref{Sfig3}e, the contrast in signal between the two trap-state for the \Ex transition gently fades as the green power is increased. We assume a drift proportional to the applied non-resonant power and retrieve the proportionality constant (Fig.\,\ref{Sfig3}f), verifying that the linear assumption is valid as the fit closely reproduces the data. This effect can then be added to the model for population evolution developed above and reproduces perfectly the droop in RF signal.

\subsection*{Resonant power: Laser background and ground state population}
\begin{figure*}
    \includegraphics[width=1\textwidth]{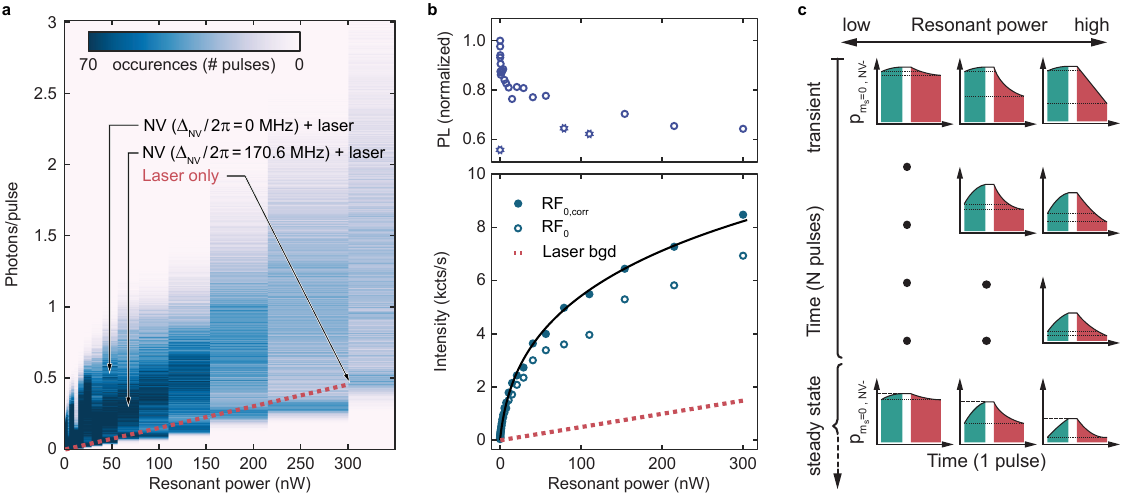}
    \caption{Resonant saturation and background correction. \textbf{a} Histogram of the average number of collected photons per pulse, binned over sequences of 600 pulses. The histogram shows three sub-distributions with centers evolving differently as a function of the applied resonant power. While two sub-distributions (corresponding to the \NV being driven on-resonance and \unit[170.6]{MHz} blue-detuned due to the influence of the nearby charge trap) exhibit a saturation behavior, the third one increases perfectly linearly with laser power, allowing us to fit and retrieve the laser background (red dashed line). \textbf{b} Top panel: Evolution of the photoluminescence intensity (during the green repump pulse) with increasing resonant laser power. The PL signal, normalized to its maximum value, is proportional to charge state and to a lesser extend spin state population. Star-shaped symbols represent measurements showing evidences of increased environmental mechanical vibration levels. As a consequence lower NRF and RF intensities were recorded. Bottom panel: the raw peak RF signal intensities (open teal circles) can be corrected first for the residual laser background (dashed red line) and second for the decreased repumping efficiency with increased resonant power (see panel \textbf{c}) using the normalized NRF signal.}
    \label{Sfig4}
\end{figure*}

Fig.\,4b of the main manuscript shows the corrected $RF_\mathrm{0,corr}$ rates, described as the resonance fluorescence rates at the start of the probing resonant pulse (i.e.\ before spin-shelving or ionization deplete the signal) corrected for resonant laser leakage and the loss of ground state population due to competition between spin-shelving and ionization (resonant laser) and spin-polarization and \NVo recombination (green non-resonant repump laser).

The laser background $\mathcal{B}_L$ is extracted by histogramming time-traces (such as presented in Fig.\,\ref{Sfig0}c) as a function of resonant power. For each histogram, three distributions can be extracted, with their means corresponding to three cases, from high to low $\overline{RF}(\Delta_\mathrm{NV}/2\pi=0)+\mathcal{B}_L$, $\overline{RF}(\Delta_\mathrm{NV}/2\pi= 170.6)+\mathcal{B}_L$ and $\mathcal{B}_L$. The third ``laser only'' distribution $\mathcal{B}_L$ builds up when the \NV is ionized and cannot fluoresce. While the two other distributions already hint at a saturation behaviour, $\mathcal{B}_L$ increases perfectly linearly with input power, comforting the attribution. A slope with units $\bigl(\mathrm{cts \cdot s^{-1} \cdot nW}\bigr)$ can then be extracted (red dashed line in Fig.\,\ref{Sfig4}a,b).

Fig.\,\ref{Sfig4}b shows on the top panel the recorded non-resonant photoluminescence signal gathered at the end of the repump pulses, i.e.\ after polarization and charge recovery. Already for moderate resonant power, the photoluminescence drops significantly. Panel c of the same figure outlines the physical process and why only the relative population can be retrieved: as the resonant power increases, a new pulse-to-pulse steady-state is reached, but at no point can we attribute a definite population value.

We nevertheless correct the $RF_0+\mathcal{B}_L$ data (open teal circles) with the previously determined laser background rate and the normalized PL data, yielding $RF_\mathrm{0,corr}$. A side effect of the normalization is the correction of values for three points (marked with star symbols) for which both PL and RF signal dip strongly. We verify that this drop in count rate is in fact attributable to an increase in background vibrations during these measurement runs by looking at the concurrently acquired noise spectra (derived from the Fourier transform of the data cast as time-traces) showing increased power spectral density values around the resonance frequency of the cavity's main stage.

\subsection*{Fitting of Rabi oscillations}
\begin{figure*}
    \includegraphics[width=1\textwidth]{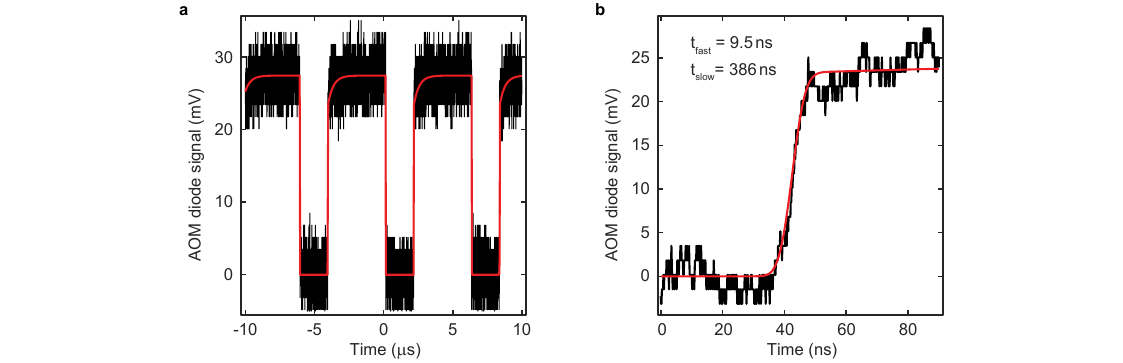}
    \caption{Acousto-optical modulator risetime. \textbf{a} Multi-period signal (black) used to fit (red) and extract the relevant timescale for the rise-time of the resonant laser pulse. \textbf{b} Close-up of the fast leading-edge of one of the optical pulses. t$_\mathrm{fast}$ corresponds to $2\sigma$ of the error function used to model the fast rise in power, while t$_\mathrm{slow}$ is the characteristic time of the inverted exponential characterizing the slower part of the rise.}
    \label{Sfig5}
\end{figure*}

The model used to fit the data in Fig.\,4c of the main text is entirely based on solving the optical Bloch equations numerically, with a temporal evolution of the driving field retrieved from the rise-time of our resonant red acousto-optic modulator (AOM) as shown in in Fig.\,\ref{Sfig5}. It is noted that this rise-time, while particularly fast for such a device, still contributes greatly to the reduction of the contrast.

Charge noise is included as described previously, by integrating the expected response of the two-level system for the extent of detunings given by previously characterized $\Gamma_\mathrm{ext}$. The shelving observed at long time-scales (see Fig.\,\ref{Sfig1}d,e) is taken into account with an ad-hoc exponential term, but is virtually imperceptible for the \Ex transition. We decided to keep the full signal in Fig.\,4c, including laser background and \A fluorescence. Both are computed and added to the Bloch equation model thanks to previous analysis: background laser rate and AOM rise-time for the laser leakage, and the amplitude and decay time measured in Fig.\,\ref{Sfig1}d for the \A fluorescence. We note that the possibility that our observations result from coherent driving of \A is vanishingly small, as the large detuning would result in fast oscillations which would be washed out by the AOM rise time (verified by our model). In fact, the same smearing effect is already taking place for our control experiment at detuning \unit[$\Delta_\mathrm{NV} = 2\pi \cdot 170.62$]{MHz}. We do not fit the latter as the exact \A contribution is unknown. However, running our model for such a detuning and without any \A contribution results in a curve reproducing the data well, downshifted by only a few percent from the measurement points and showing no rapid oscillations (as expected).

\bibliography{biblio_short}